\documentclass[%
 aip,
 jap,
 amsmath,amssymb,
reprint,%
]{revtex4-1}

\usepackage{graphicx}
\usepackage{dcolumn}
\usepackage{bm}

\begin{document}

\preprint{AIP/JAP-1}

\title[Hot carriers in a bipolar graphene]{Hot carriers in a bipolar graphene}

\author{O.G. Balev}
\affiliation{Departamento de Fisica, Universidade Federal do Amazonas,
Manaus, 69077-000, Brazil}
\author{F.T. Vasko}
\email{ftvasko@yahoo.com}
\affiliation{Institute of Semiconductor Physics, NAS of Ukraine,
Pr. Nauki 41, Kiev, 03028, Ukraine}
\date{\today}

\begin{abstract}
Hot carriers in a doped graphene under dc electric field is described taking
into account the intraband energy relaxation due to acoustic phonon scattering
and the interband generation-recombination transitions caused by thermal radiation.
The consideration is performed for the case when the intercarrier scattering 
effectively establishes the quasiequilibrium electron-hole distributions, with 
effective temperature and concentrations of carriers. The concentration and energy 
balance equations are solved taking into account an interplay between weak energy 
relaxation and generation-recombination processes. The nonlinear conductivity is 
calculated for the momentum relaxation caused by the elastic scattering. The 
current-voltage characteristics, and the transition between bipolar and monopolar 
regimes of conductivity are obtained and analyzed, for different temperatures 
and gate voltages.
\end{abstract}



\maketitle

\section{Introduction}

Study of the heated under dc electric field nonequilibrium carriers in graphene is 
stimulated by attempts of realization of efficient field-effect transistor, see 
\cite{1} and Refs. therein. Essential heating also takes place for the cleaning of 
graphene by strong pulses of applied current \cite{2} and in recent measurements, 
\cite{2a} conducted in electric fields $E\geq$1 kV/cm. In addition, excited by 
ultrafast interband pumping electron-hole pairs are actively studied \cite{3}. In 
connection with such experiments theoretical studies of energy relaxation, both 
analytical \cite{4} and numerical \cite{5}, were conducted as well as studies of 
relaxation dynamics after ultrafast photoexcitation \cite{6}. Recently, investigation 
of carriers heating under dc electric field applied to an intrinsic graphene has 
been given in \cite{7}. Main peculiarities of  the processes of heating in graphene 
(gapless and massless semiconductor, which is described by the Weyl-Wallace model 
\cite{8}) are related with an interplay between quasi-elastic energy relaxation on 
acoustic phonons and generation-recombination processes, that are effective nearby 
the cross point of energy spectrum. In a gated graphene, when monopolar regime of 
transport is realized, contribution of generation-recombination processes is suppressed. 
Effect of the gate voltage on the processes of carriers 
heating and on current-voltage characteristics of a gated graphene (i.e., a study
of the model of field transistor located on graphene sheet of large dimensions) 
have not been considered so far.

In this paper a heating of carriers in doped graphene is analyzed theoretically, for 
the case when difference between the densities of electrons and holes is controlled by 
a gate voltage. Present treatment is limited by the region of energies below the energy 
of optical phonon ($\omega_{o} \approx 0.2$eV), when it is necessary to take into 
account the Joule heating produced by dc current, induced by electric field $\textbf{E}$, 
as well as following processes of relaxation: a) elastic relaxation of momentum due 
to elastic scattering on structural disorder \cite{9}, which is the most effective
mechanism of scattering so that anisotropy of carriers distributions is small; b) 
quasi-elastic relaxation of energy on acoustic phonons; and c) generation-recombination 
processes for interband transitions caused by thermal radiation (relaxation of energy 
and concentrations were analyzed for photoexcitation in \cite{10}). Here we consider 
the regime when the interparticle Coulomb scattering dominates over the processes 
of relaxation of the energies and the concentrations, so that the carriers are described 
by quasiequilibrium distributions with effective temperature and nonequilibrium 
concentrations of electrons and holes. These parameters are defined from the balance 
equations of the concentrations and the total energy, where the areal charge density 
of graphene sheet, given by the difference of the electron and the hole densities, 
is controlled by gate voltage. Besides determination of the average energies of carriers 
and their concentrations, nonlinear current-voltage characteristics are also studied 
below.

The analysis performed below is organized as follows. The basic equations, describing 
heating of carriers in gated graphene are evaluated in Sec. II. Results of calculations, 
including the parameters of quasiequilibrium distributions, nonequilibrium concentrations 
and current-voltage characteristics as a function of applied field, gate voltage, and temperature are presented in Sec. III. Discussion of the assumptions used and concluding remarks are given in Sec. IV. In Appendix the collision integrals used in present treatment 
are specified.

\section{Basic equations}
Our description of the heating of bipolar plasma in graphene under dc electric field is 
based on quasiclassical kinetic equations that take into account the scattering 
mechanisms listed above. Here we assume that $\nu_{m}\gg \nu_{cc} \gg \nu_{qe,r}$, 
where $\nu_{m}$, $\nu_{cc}$, and $\nu_{qe}$ ($\nu_{r}$) are the relaxation frequencies of 
momentum, intercarrier scattering, and energy (concentration), respectively. Due to 
dominance of the relaxation of momentum, the weak anisotropic contributions to the 
electron ($e$) and the hole ($h$) distributions are given as \cite{9}
\begin{equation}
\Delta f_{k{\bf p}}=\pm\frac{(e{\bf E}\cdot{\bf p})}{p\nu_p^{(m)}}
\left( -\frac{df_{kp}}{dp}\right) .
\label{1}
\end{equation}
Here $k=e$ ($k=h$) corresponds the upper (lower) sign and the frequency of momentum 
relaxation $\nu_{p}^{(m)}=v_{m}p\Psi (pl_{c}/\hbar )/\hbar$ is expressed by means of 
the characteristic velocity $v_{m}$ (that defines efficiency of scattering), and of 
the truncation factor for long-range scattering \cite{9} $\Psi(x)=x^{-2}\exp(-x^{2})I_{1}(x^{2})$. 
Here the correlation length of static disorder, $l_{c}$, and the 
modified Bessel function of the first kind, $I_{1}(x)$, are used. 
The nonequilibrium isotropic distributions $f_{kp}$ are defined by kinetic equations
\begin{equation}
\pm\overline{e{\bf E}\cdot\frac{\partial\Delta f_{kp}}{\partial{\bf p}}}
= J_{qe}\left( f|kp\right)+J_r\left( f|p\right)+J_{cc}\left( f|kp\right) ,
\label{2}
\end{equation}
where an overline stands for the averaging over the angle in ${\bf p}$-plane. The 
collision integrals in the right hand side of Eq. (\ref{2}) are given in Appendix.

Obtained from Eq. (\ref{2}) distributions $f_{kp}$ define concentrations of electrons, 
$n_{e}$, and holes, $n_{h}$, as follows
\begin{equation}
n_k=4\int\frac{d{\bf p}}{(2\pi \hbar )^2}f_{kp} ,
\label{3}
\end{equation}
where the factor $4$ takes into account degeneration over the spin and the valleys.
The current density, $\textbf{I}$, is given as
\begin{equation}
{\bf I} = 4e\int\frac{d{\bf p}}{(2\pi\hbar )^2}{\bf v}_{\bf p}\left( \Delta
f_{e{\bf p}}-\Delta f_{h{\bf p}} \right) ,
\label{4}
\end{equation}
where ${\bf v}_{\bf p}=v_W{\bf p}/p$ is the velocity of particle with momentum $\textbf{p}$,
written using the characteristic velocity $v_{W} \approx 10^{8}$cm/s.  The sheet charge is
defined by the difference of concentrations $n_{e}-n_{h} \equiv \Delta n_{s}$ that is 
controlled by the gate voltage $V_{g}$ applied to the back gate, placed at a distance $d$ 
from the graphene sheet. Considering such structure as plane capacitor, we obtain relation 
between $\Delta n_{s}$ and $V_{g}$ as follows: $\Delta n_{s}=\epsilon V_{g}/(4 \pi |e|d)$, 
where $\epsilon \approx 3$ is dielectric constant of SiO$_{2}$ substrate.

Because of predominance of the intercarrier scattering in Eq. (\ref{2}), the symmetric 
distributions $f_{kp}$ must satisfy the following conditions
\begin{eqnarray}
f_{kp+\Delta p}f_{k_1 p^{\prime}-\Delta p}(1-f_{kp})(1-f_{k_1 p^{\prime}}) \\
= f_{kp}f_{k_1 p^{\prime}}(1-f_{kp+\Delta p})(1-f_{k_1 p^{\prime}-\Delta p}) , \nonumber
\label{5}
\end{eqnarray}
which nullify the factor in integrand of the intercarrier collision integral (\ref{A3}). 
Hence $J_{cc}$ imposes the quasiequlibrium distributions
\begin{equation}
\widetilde{f}_{kp}=\left[\exp\left(\frac{v_W p-\mu_k}{T_c}\right)+1\right]^{-1} ,
\label{6}
\end{equation}
where $T_{c}$ is the effective temperature of carriers, and $\mu_{k}$ are the electrochemical 
potentials of carriers. These three parameters are related by requirement for the area charge 
density to be constant and equal to $e \Delta n_{s}$, which gives the electroneutrality condition in the form
\begin{equation}
e\Delta n_s=4e\int\frac{d{\bf p}}{(2\pi\hbar )^2}(\widetilde{f}_{ep}
-\widetilde{f}_{hp}) .
\label{7}
\end{equation}
Further, due to conservation of concentrations of electrons and holes under the intercarrier 
scattering processes [see Eqs. (\ref{A4})] we obtain the concentration balance equation 
written through the radiative collision integral (\ref{A2})
\begin{equation}
\int\frac{d{\bf p}}{(2\pi \hbar )^2}\nu_p^{(r)} \left[ N_{2p/p_T}(1-f_{ep}-f_{hp})-f_{ep}f_{hp}
\right]=0 .
\label{8}
\end{equation}
Here $N_{z}=(e^{z}-1)^{-1}$ is the Planck distribution function 
of the equilibrium thermal radiation with temperature $T$ and we have introduced the characteristic 
thermic momentum $p_{T}=T/v_{W}$. Point out, to obtain the concentration 
balance equation (\ref{8}) we have used that: $n_{k}$ is not modified by scattering on 
acoustic phonons (because of smallness of the velocity of sound in comparison with $v_{W}$) 
and $J_{r}(f|p)$ is independent of $k$.

In a similar way, by summing Eq. (\ref{2}) over $\textbf{p}$ and $k$ with the weight 
$v_{W} p$, we obtain the energy balance equation: $P(T_c)=Q_J$. Here the energy 
losses term, $P(T_c)$, is written as follows:
\begin{equation}
P(T_c)=-4v_W\int\frac{d{\bf p}p}{(2\pi \hbar )^2}\left[\sum_{k}J_{qe}\left(\widetilde{f}
|kp\right)+2J_r\left(\widetilde{f}|p\right) \right] ,
\label{9}
\end{equation}
while the Joule heating term, $Q_{J}={\bf I} \cdot {\bf E}$, is given as
\begin{equation}
Q_J=-4e v_W\int\frac{d{\bf p}p}{(2\pi \hbar )^2}\overline{{\bf E}\cdot\left(
\frac{\partial\Delta f_{ep}}{\partial{\bf p}}-\frac{\partial\Delta f_{hp}}
{\partial {\bf p}} \right)} .
\label{10}
\end{equation}
Using the integration by parts, one can rewrite Eq. (\ref{10}) as $Q_J=\sigma E^2$,
where the nonlinear conductivity $\sigma$ is introduced by the relation 
${\bf I}=\sigma {\bf E}$. Under substitution Eq. (1) into (4) we obtain the conductivity
\begin{equation}
\sigma =\sigma_0\left[ F_{p=0}+\frac{1}{2}\int\limits_0^\infty
dp F_p\frac{d}{dp}\Psi\left(\frac{pl_c}{\hbar}\right)^{-1}\right] ,
\label{11}
\end{equation}
where $F_p\equiv f_{ep}+f_{hp}$ and $\sigma_0 =(2v_W /v_d)e^2/\pi\hbar$ is the 
linear response conductivity for the short-range scattering in intrinsic graphene.

Thus, the parameters of the distributions Eq. (\ref{6}) are defined by transcendent equations
(\ref{7})-(\ref{11}). The nonequilibrium concentrations $n_{k}$ and the current density
$\textbf{I}$ are determined by Eq. (\ref{3}) and Eq. (\ref{4}), respectively.

\section{Results}
Below we discuss the solutions of Eqs. (\ref{7})-(\ref{11}) as well as the concentrations
and the current-voltage characteristics versus the field $E$, the gate voltage $V_{g}$,
and the temperature $T$. Calculations are performed at temperature interval $T=$77 - 300 K 
for the graphene on SiO$_{2}$ substrate of width $d\simeq 3\times 10^{-5}$ cm and for the correlation length $l_c\simeq$10 nm.

\subsection{Quasi-equilibrium distribution}
The quasiequilibrium distributions given by Eq. (\ref{6}) are determined from the balance
equations, (\ref{7})-(\ref{10}), where it is convenient to use the dimensionless
momentum, $x=v_{W}p/T_{c}$, such that $\widetilde{f}_{kx}=[\exp (x-\mu_{k} /T_c)+1]^{-1}$.
Then the charge neutrality condition (\ref{7}) obtains the form
\begin{equation}
\int_0^{\infty}dx x\left(\widetilde{f}_{ex}-\widetilde{f}_{hx}\right) =\frac{\pi}{2}
\left(\frac{\hbar v_W}{T_c}\right)^2\Delta n_s .
\label{12}
\end{equation}
The concentration balance equation (\ref{8}) is  given as
\begin{equation}
\int_0^{\infty}dx x^{2} \widetilde{f}_{ex} \widetilde{f}_{hx}
\left\{\frac{e^{2x-(\mu_{e}+\mu_{h})/T_{c}}-1}{e^{2xT_{c}/T}-1}-1\right\} =0 ,
\label{13}
\end{equation}
and it is not dependent explicitly on the field $E$ or gate voltage $V_{g}$.
Multiplying the energy balance equation $P(T_c)=Q_J$ by $\pi\hbar^{3}v_{W}^{3}
/(2v_{qe} T_{c}^{4})$ we obtain its dimensionless form 
$\widetilde{P}(T_c)=Q_{E}$. Where from Eqs. (\ref{9}) and (\ref{10}) it follow the 
dimensionless energy losses
\begin{eqnarray}
\widetilde{P}(T_c) =\frac{T_{c}-T}{T}\sum_{k=e,h}\int_0^{\infty} dx x^{4} e^{x-\mu_{k}/T_c}
      \widetilde{f}_{kx}^{2} ~~~~ \nonumber \\  
-2\Gamma \int_0^{\infty}dx x^{3}\widetilde{f}_{ex} \widetilde{f}_{hx}
\left\{\frac{e^{2x-(\mu_{e}+\mu_{h})/T_{c}}-1}{e^{2xT_{c}/T}-1}-1\right\} ,  
\label{14}
\end{eqnarray}
and the dimensionless Joule heating
\begin{eqnarray}
Q_E=\left(\frac{v_W p_E}{T_{c}}\right)^4\left[\widetilde{f}_{ex=0}+\widetilde{f}_{hx=0}
\right. \nonumber \\
\left. -\frac{\eta_c}{2}\int_0^\infty dx\left(\widetilde{f}_{ex}+\widetilde{f}_{hx}\right)
\frac{\Psi '(\eta_{c} x)}{\Psi (\eta_{c} x)^{2}}\right] .
\label{15}
\end{eqnarray}
Here the characteristic momentum $p_{E}$ is introduced as  $p_{E}^{4}=(eE\hbar)^{2}
/v_{qe}v_d$ and $\eta_c =T_cl_c/\hbar v_W$.

For heavily doped case electrons are degenerated and holes are nondegenerated (or vice
versa). Then the electron quasi-Fermi energy, $\mu_{e}$, 
is given from Eq. (\ref{12}) as $\mu_{e}\simeq\sqrt{\pi n_{e}} \hbar v_{W}$
which is weakly dependent of temperature.
Further, equation that defines $T_{c}$ (e.g., as function of $E$, for given $V_{g}$ and 
$T$) is obtained from Eq. (\ref{15}) as
\begin{equation}
T_c\simeq T\left[1+\frac{\left( v_W p_E /\mu _e \right)^4}{2 \Psi \left(\mu _e l_c
/\hbar v_W\right)} \right] .
\label{17}
\end{equation}
Point out, the condition of heavily doping (in particular, $n_{e}\gg n_{h}$) we can 
rewrite as $(\mu_{e}/T_{c})^{2} \gg 1$. 

\begin{figure}[ht]
\begin{center}
\includegraphics[scale=0.4]{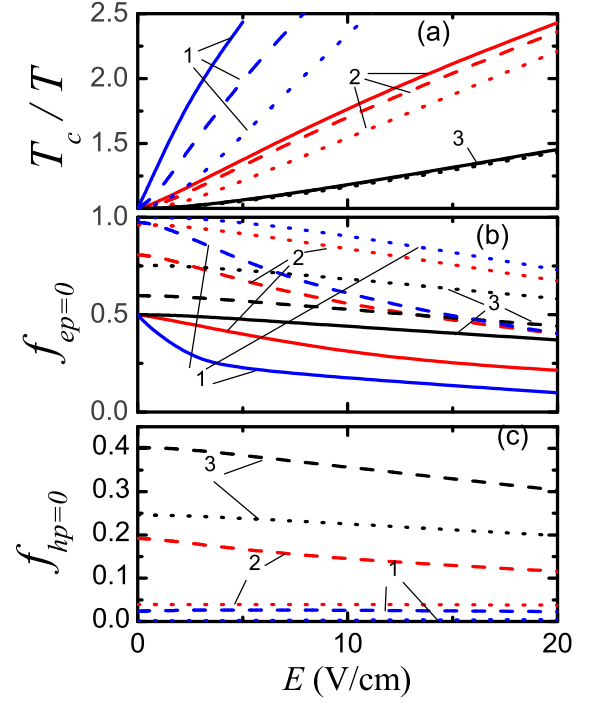}
\end{center}
\addvspace{-2 cm}
\caption{(Color online) Effective temperature $T_c/T$ (a) and maximal distribution of 
electrons $f_{ep=0}$ (b) and holes $f_{hp=0}$ (c) versus electric field for $T=$77 K (1), 
150 K (2), and 300 K (3). Solid, dashed, and dotted curves correspond to $V_g=$0 V, 
1 V, and 3 V, respectively.}
\end{figure}

\begin{figure}[ht]
\begin{center}
\includegraphics[scale=0.4]{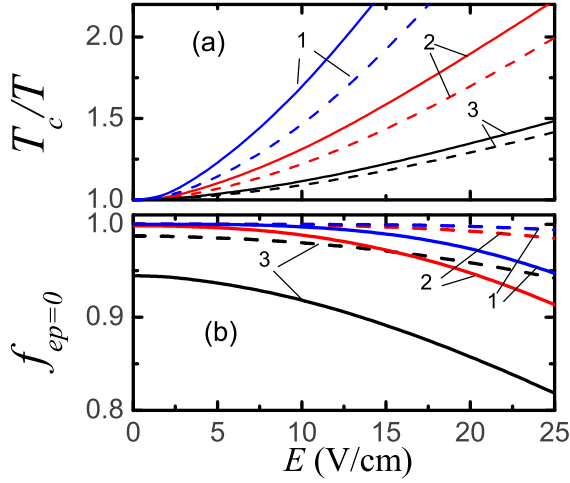}
\end{center}
\addvspace{-4 cm}
\caption{(Color online) Effective temperature $T_c/T$ (a) and maximal distribution of 
electrons $f_{ep=0}$ (b) versus electric field for T =77 K (1), 150 K (2), and 300 K (3). 
Solid and dashed curves correspond to $V_g=$10 V and 20 V, respectively. }
\end{figure}

In Figs. 1a, 1b, and 1c we plot the effective temperature $T_{c}/T$ and the
maximal distributions, $f_{ep=0}$ and $f_{hp=0}$, for different doping levels
at $V_g$=0 V, 1 V, and 3 V [notice, in Fig. 1(c) the solid curves, pertinent to 
$V_{g}$=0 V, are omitted as they coincide with the solid curves of Fig. 1b].
It is seen from Fig. 1a that, for given $V_g$, the relative increase of the 
effective temperature of carriers, $T_{c}/T$, with growing $E$ becomes smaller for 
larger $T$. 
Fig. 1b shows that the decrease of $f_{ep=0}$ with growing $E$ becomes, at given $T$, 
slower for larger $V_g$. From Figs. 1c and 1b it is seen that $f_{hp=0}$, for given 
$V_g$, quickly decreases as $T$ grows. 
For heavily doped case, at $V_{g}=$10 V and 20 V, in Figs. 2a and 2b we plot 
$T_{c}/T$ and $f_{ep=0}$ as functions of $E$. For these gate voltages,
the hole concentrations are small, so that $f_{hp=0}<0.1$ and we do not plot the
dependencies $f_{hp}$ here. The dependencies $T_c/T$ and $f_{ep=0}$ on $T$, $V_g$, and $E$
now are similar to those in the low-doping region. 

\begin{figure}[ht]
\begin{center}
\includegraphics[scale=0.4]{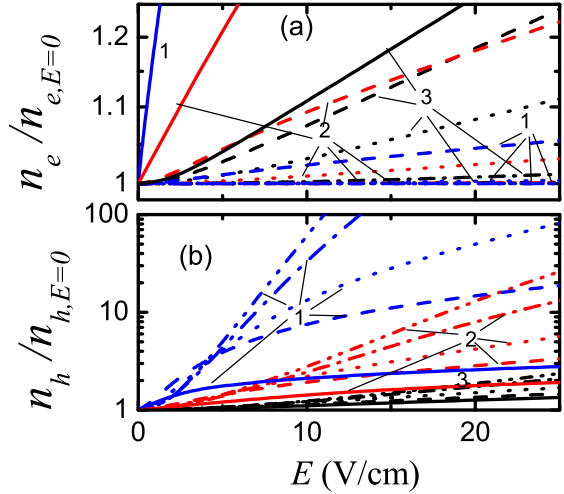}
\end{center}
\addvspace{-3.5 cm}
\caption{(Color online) Electron (a) and hole (b) concentrations (normalized to their
values at $E=0$) versus electric field at $T=$77 K (1), 150 K (2), and 300 K (3). The solid, 
the dash, the dot, the dash-dot, and the dash-dot-dot curves correspond to $V_g=$0 V, 1 V, 3 V, 10 V, and 20 V, respectively. }
\end{figure} 

\subsection{Carrier concentrations}
Using solutions of the balance Eqs. (\ref{12})-(\ref{15}) in Eq. (\ref{3}) we
plot in Fig. 3a dimensionless concentrations of electrons, $n_{e}/n_{e,E=0}$, 
and in Fig. 3b of holes, $n_{h}/n_{h,E=0}$. Here $n_{e,E=0}$ and $n_{h,E=0}$ are 
equilibrium concentration of electrons and holes for $E=0$ and these concentrations 
are strongly dependent on $V_{g}$ and $T$. \cite{11} For small concentrations (solid 
and dashed curves in Fig. 3a) of electrons (the main carriers) their concentration 
is increased on tens of percents for growing $E$. However, for $V_{g} \geq 10$V 
the concentration of main carriers becomes very weakly growing function of $E$
even at $T=300$K. At the same time, the density of the minor carriers (holes) 
according to Fig. 3b can be enlarged by tens of times, especially at low temperatures.  

\begin{figure}[ht]
\begin{center}
\includegraphics[scale=0.4]{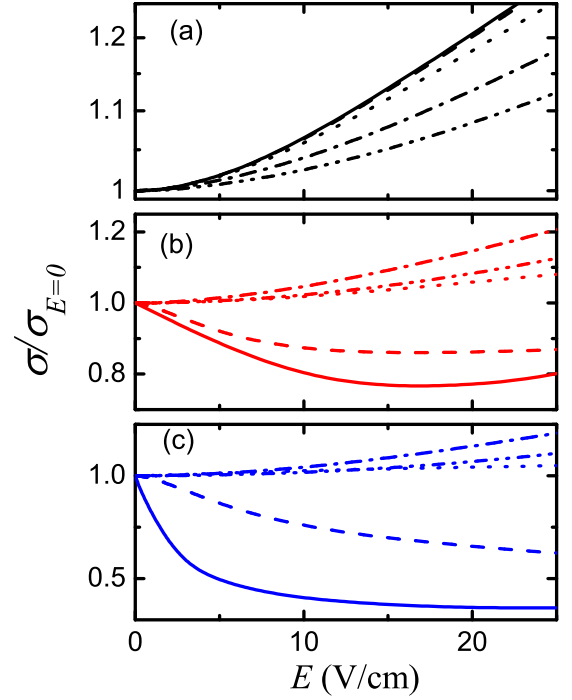}
\end{center}
\addvspace{-1 cm}
\caption{(Color online) Nonlinear conductivity, normalized to its value at $E=0$, 
versus electric field for: $T=$300 K (a), 150 K (b) and 77 K (c). The solid, 
the dash, the dot, the dash-dot, and the dash-dot-dot curves correspond to 
$V_g=$ 0 V, 1 V, 3 V, 10 V, and 20 V, respectively. }
\end{figure}


\subsection{Current-voltage characteristics}
In Fig. 4, we plot the normalized conductivity $\sigma/\sigma_{E=0}$ versus 
field $E$ for different levels of doping, dependent on gate voltage varied
between $V_{g}=0$ and $V_{g}=$20 V and on different temperatures. Point out 
$\sigma_{E=0}$ for each curve in Figs. 4 a-c has specific value and 
$\sigma_{E=0}/\sigma_{0}$ increases essentially when $V_g$ grows. \cite{12}
The field dependency of the conductivity is determined by a competition  between the first
and the second terms in Eq. (\ref{11}). Indeed, as $F_{p=0}$ is a decreasing function 
of $E$ the integral (second) contribution, for $\mu_{e}l_{c}/\hbar v_{W } \agt 1$ or/and 
$T_{c}l_{c}/\hbar v_{W} \agt 1$, is either much slower decreasing function of $E$ 
or even an increasing function of $E$. In particular, for $T_{c}l_{c}/\hbar v_{W} \agt 1$ 
and $\mu_{e}/T_{c} \gg 1$, the second contribution is the integral of the product of 
$F_{p}$ (that is $\approx F_{p=0}$ for $\mu_{e}/v_{W} \geq p \geq 0$ and drops rapidly 
for $p> \mu_{e}/v_{W}$) with the other factor  that is a large and growing function of 
$p$, $\propto p^{2}$, in actual region of $p$, $T_{c}/v_{W} \alt p \leq \mu_{e}/v_{W}$.
So for low temperatures the contribution $F_{p=0}$ (see Fig. 4 b and c) is essential and 
leads to the appearance of a minimum in the field dependence
of $\sigma$ for small concentrations, for $V_{g}<$ 3V. With the growth of electric field 
or the concentration (for $V_g>$3 V) the main contribution to the conductivity comes from
the integral term in Eq. (\ref{11}) and now $\sigma$ is monotinically growing function of 
$E$, cf. Fig. 4 a, b, and c. Indeed, pertinent current-voltage characteristics show 
transition from a sublinear dependence to a superlinear one, as the concentration 
or/and the temperature grow. For $V_g\geq$3 V (and at $T=$300 K for any $V_{g}$, or 
concentration) a superlinear current-voltage characteristic is realized.  
     
\section{Conclusions}
In the present work, we investigate the effect of doping on carrier heating in a gated 
graphene based on the balance equation approach. The effective temperature and concentrations
of carriers are studied as functions of the thermostat temperature, the gate voltage, and the applied
dc field. Pertinent current-voltage characteristics  show the transition from the regime of heating for intrinsic 
bipolar conductivity to the regime of heating for the conductivity of strongly degenerated carriers. 
In the latter case the heating is well described by simple analytical formulas, see (16).

Next, we list and discuss the assumptions used in the calculations performed.
Here we have examined the heating of carriers in the low energy region only, when
the optical phonon emission is not essential and the radiative-induced direct 
interband transitions are assumed to be the main generation-recombination mechanism.
Point out that for much larger electric fields (for $E\gg$0.1 kV/cm, see \cite{2a}) 
present approach will not be valid as the optical phonon contribution is not taken into account,
however, the influence of optical phonons is smaller for a higher level of the doping (gate voltage).
We have also restricted 
ourselves by the study of limiting case when the intercarrier Coulomb scattering 
is dominating. As shown in \cite{7} for the intrinsic graphene case, this is
an adequate approach. Possible contribution of other generation-recombination 
mechanisms (note, that the Auger-processes are forbidden due to the symmetry of 
electron-hole states \cite{13}) require an additional investigation. Further, we 
have assumed that the heat removal is sufficiently effective (this point becomes 
insignificant in the case of short  electric  pulses when the thermostat is not 
overheated). At last, to describe the momentum relaxation we take into account 
only the statical disorder scattering using the phenomenological model of Ref. 
10 (which is in good agreement with the experimental data; the microscopic
mechanisms of scattering are still unclear \cite{14}). The listed assumptions 
should not essentially modify the peculiarities of the heating mechanisms, in comparison 
with a study that will treat relaxation processes in more details; in addition,
here many limitations are imposed also because of the lack of data on graphene. 

To conclude, essential heating of carriers is shown in present study. It will
define different parameters of possible graphene-based devices. In addition, investigation
of hot carriers (even within the considered limited region of parameters) gives important 
information about mechanisms of relaxation and recombination in graphene. 
Therefore, further experimental and theoretical (including numerical modeling)
study of a heating in graphene is very opportunely now.

\appendix*
\section{Collision integrals}

Below we present the collision integrals $J_{qe}$, $J_{r}$ and $J_{cc}$ (deduced in
\cite{9,10} and \cite{15}, respectively) that have been used to obtain the balance equations.
For quasi-elastic scattering on acoustic phonons the energy relaxation is described by
the Fokker-Planck nonlinear differential form
\begin{equation}
J_{qe}\left( f|kp\right)=\frac{\nu _p^{(qe)}}{p^2}\frac{d}{dp}\left\{ p^4 \left[
\frac{df_{kp}}{dp}+\frac{f_{kp}(1-f_{kp})}{p_T}\right] \right\} .
\label{A1}
\end{equation}
Here $k=e,h$ and it is introduced the frequency of quasi-elastic relaxation
$\nu_{p}^{(qe)}=v_{qe} p/\hbar$ that contains the characteristic velocity 
$v_{qe}\propto T$, where for typical parameters of graphene we have $v_{qe} \approx 
51.4$ cm/s at $T=300$K. The thermal-radiation-induced interband transitions are being 
depicted by the collision integrals
\begin{equation}
J_r \left( f|p\right)=\nu_p^{(r)}\left[ N_{2p/p_T}(1-f_{ep}-f_{hp})-f_{ep}f_{hp}
\right] ,
\label{A2}
\end{equation}
that are equal for electrons and holes. In Eq. (\ref{A2}) it is used the frequency 
of radiative relaxation $\nu_{p}^{(r)}=v_{r} p/\hbar$, where the characteristic velocity 
$v_{r} \approx 41.6$cm/s for graphene imbedded between SiO$_{2}$ substrate and cover layer.

As Auger-processes are forbidden because of the electron-hole symmetry of the bands in 
graphene \cite{13}, the intercarrier collision integral receives the form
\begin{eqnarray}
J_{cc}\left( f|kp\right) =\sum\limits_{k_1=k,k'}\int\limits_0^\infty d p'
\int\limits_{-\infty}^\infty d\Delta p W_{kk_1} ( p p'|\Delta p) \nonumber \\
\times\left[ f_{k p+\Delta p}f_{k_1 p'-\Delta p}(1-f_{k p})(1-f_{k_1 p'}) \right. \\
\left.-f_{k p}f_{k_1 p'}(1-f_{k p+\Delta p})(1-f_{k_1 p'-\Delta p})\right]  , \nonumber
\label{A3}
\end{eqnarray}
where it describes the transitions due to carrier-carrier scattering from the initial
states ($k \textbf{p}, k^{\prime} \textbf{p}^{\prime}$) to the states ($k \textbf{p}+
\Delta \textbf{p},k^{\prime} \textbf{p}^{\prime}-\Delta \textbf{p}$).
Here $v_{W}|\Delta \textbf{p}|$ determines transference of the energy for scattering 
and $k=k'$ corresponds to electron-electron or hole-hole transitions whereas the channel 
$k \neq k'$ describes scattering of electrons on holes. Also in Eq. (\ref{A3}) it is 
carried out average over the angle and as a result the probability of transition $W_{kk'} 
(pp'|\Delta p)$ is dependent from $|\textbf{p}|$, $|\textbf{p}'|$ and from the 
transference of the energy $v_{W}\Delta p$. Deduction of the concentration (\ref{8}) 
and the energy (\ref{9}) balance equations, needed, along with Eq. (\ref{7}),
to obtain the parameters of quasiequilibrium distributions Eq. (\ref{6}), is based on 
the property of conservation as of the concentration so of the energy density for 
interparticle scattering as
\begin{equation}
\frac{4}{{L^2 }}\sum\limits_{\bf p} {J_{cc} \left( {f|kp} \right) = 0,} ~~~
\frac{{4v_W }}{{L^2 }}\sum\limits_{\bf p} {pJ_{cc} \left( {f|kp} \right) = 0.}
\label{A4}
\end{equation}
These conditions straightforwardly follow from Eq. (\ref{A3}) if take into account the 
symmetry of the probability of transitions under the interchange of scattering channels 
$k, \textbf{p}$ and $k',\textbf{p}'$.

\begin{acknowledgments}
This work of O. G. B. was supported by Brazilian FAPEAM  (Funda\c{c}\~{a}o de Amparo \`{a} Pesquisa do Estado do
Amazonas) Grant.
\end{acknowledgments}


\begin{thebibliography}{12}
\bibitem{1}
M. C. Lemme, Solid State Phenomena, {\bf 156-158}, 499 (2010).

\bibitem{2}
J. Moser, A. Barreiro, and A. Bachtold, Appl. Phys. Lett. {\bf 91}, 163513 (2007).

\bibitem{2a}
I. Meric, M. Y. Han, A. F. Yang, B. Ozyilmaz, P. Kim, and K. L. Shepard, Nature
Nanotech. {\bf 3}, 654 (2008); A. Barreiro, M. Lazzeri, J. Moser, F. Mauri,
and A. Bachtold, Phys. Rev. Lett. {\bf 103}, 076601 (2009).

\bibitem{3}
J. M. Dawlaty, S. Shivaraman, M. Chandrashekhar, F. Rana, and M. G. Spencer
Appl. Phys. Lett. {\bf 92}, 042116 (2008).
D. Sun, Z.-K. Wu, C. Divin, X. Li, C. Berger, W. A. de Heer, P. N.First,
and T. B. Norris, Phys. Rev. Lett. {\bf 101}, 157402 (2008).
P. A. George, J. Strait, J. Dawlaty, S. Shivaraman, Mvs. Chandrashekhar, F. Rana,
and M. G. Spencer, Nanoletters {\bf 8}, 4248 (2008).
R. W. Newson, J. Dean, B. Schmidt, and H. M. van Driel, Opt. Exp. 17, 2326 (2009).

\bibitem{4}
E. H. Hwang, B.Y.-K. Hu, and S. Das Sarma, Phys. Rev. B {\bf 76},
115434 (2007); W.-K. Tse, S. Das Sarma, Phys. Rev. B {\bf 79}, 235406 (2009);
R. Bistritzer and A. H. MacDonald, Phys. Rev. Lett. {\bf 102}, 206410 (2009).

\bibitem{5}
A. Akturk and N. Goldsman, J. Appl. Phys. {\bf 103}, 053702 (2008);
R. S. Shishir and D. K. Ferry, J. Phys.: Condens. Matter, {\bf 21}, 344201 (2009).

\bibitem{6}
S. Butscher, F. Milde, M. Hirtschulz, E. Malic, and A. Knorr, Appl. Phys. Lett.
{\bf 91}, 203103 (2007); F. Rana, P. A. George, J. H. Strait, J. Dawlaty, 
S. Shivaraman, Mvs Chandrashekhar, and M. G. Spencer, Phys. Rev. B {\bf 79}, 
115447 (2009); P.N. Romanets and F.T. Vasko, Phys. Rev. B {\bf 81}, 085421  (2010).

\bibitem{7}
O.G. Balev, F.T. Vasko, and V. Ryzhii, Phys. Rev. B {\bf 79}, 165432  (2009).

\bibitem{8}
E.M. Lifshitz, L.P. Pitaevskii, and V.B. Berestetskii, Quantum Electrodynamics
(Butterworth-Heinemann, 1982); P.R. Wallace,  Phys. Rev. {\bf 71}, 622 (1947).

\bibitem{9}
F.T. Vasko and V. Ryzhii, Phys. Rev. B {\bf 76}, 233404 (2007).

\bibitem{10}
A. Satou, F.T. Vasko, and V. Ryzhii, Phys. Rev. B {\bf 78}, 115431 (2008);
F.T. Vasko and V. Ryzhii, Phys. Rev. B {\bf 77}, 195433 (2008).

\bibitem{11} 
For $V_{g}=$0, 1, 3, 10, and 20 V one obtains $n_{e,E=0}\approx 8.14\times 10^{10}$,
$1.13 \times 10^{11}$, $1.95 \times 10^{11}$, $5.56 \times 10^{11}$, and $1.10 \times 10^{12}$ cm$^{-2}$
at $T=$300 K. The correspondent hole concentrations are $n_{h,E=0}\approx 8.14 \times 10^{10}$,
$5.78 \times 10^{10}$, $3.00 \times 10^{10}$, $5.71 \times 10^{9}$, and $1.26 \times 10^{9}$ cm$^{-2}$,
i.e., the number of holes is negligible at $V_g\geq$10 V. 
At $T=$150 and 77 K, essentially lower $n_{e,E=0}$, $n_{h,E=0}$ are obtaned for $V_{g} \leq 1$ V;
for $V_{g} \geq 3$ V  $n_{e,E=0}$ tends to an independent from $T$ value $\propto V_{g}$ and $n_{h,E=0}$  
tends to exponentially small value  $\propto \exp(-b_{0}\sqrt{V_{g}}/T)$, where $b_{0}$ is a 
coefficient. 

\bibitem{12} 
For $V_{g}=$0, 1, 3, 10, and 20 V one obtains $\sigma_{E=0}/\sigma_{0}=$1.91, 1.97, 
2.34, 5.07, and 10.79 at $T=$300 K (Fig. 4a), $\sigma_{E=0}/\sigma_{0}=$1.15, 1.26, 
1.73, 4.25V, and 9.57 at $T=$150 K (Fig. 4b), and $\sigma_{E=0}/\sigma_{0}=$1.035,
1.19, 1.65, 4.02, and 9.24 at $T=$77 K (Fig. 4c).

\bibitem{13} 
M.S. Foster and I.L. Aleiner, Phys. Rev. B {\bf 79}, 085415  (2009).

\bibitem{14}
 L.A. Ponomarenko, R. Yang, T.M. Mohiuddin, M.I. Katsnelson, K.S. Novoselov, S.V. Morozov, A.A. Zhukov,
F. Schedin, E.W. Hill,   and A. K. Geim, Phys. Rev. Lett. {\bf 102}, 206603 (2009); 
S. Adam, P.W. Brouwer, and S. Das Sarma, Phys. Rev. B {\bf 79}, 201404(R) (2009).

\bibitem{15}
L. Fritz, J. Schmalian, M. Mueller, and S. Sachdev, Phys. Rev. B
{\bf 78}, 085416 (2008); M. Mueller, L. Fritz, and S. Sachdev, \textit{ ibid.}
{\bf 78}, 115406 (2008).

\end{thebibliography}
\end{document}